\documentclass[fleqn,showpacs,nofootinbib]{revtex4}
\usepackage{amssymb,amsmath,graphicx,epsfig,bm}
\DeclareMathOperator{\tr}{Tr}

\renewcommand{\d}{\mathrm{d}}

\newcommand{\sT}{{\scriptscriptstyle T}}
\newcommand{\qs}{q \!\!\! /}
\newcommand{\ks}{k \!\!\! /} 
\newcommand{\ps}{p \!\!\! /}
\newcommand{\epss}{\varepsilon\!\!\!/}

\def\slash#1{\setbox0=\hbox{$#1$}               
        \dimen0=\wd0                            
        \setbox1=\hbox{/} \dimen1=\wd1          
        \ifdim\dimen0>\dimen1                   
        \rlap{\hbox to \dimen0{\hfil/\hfil}}    
        #1                                      
        \else              
        \rlap{\hbox to \dimen1{\hfil$#1$\hfil}} 
        /                                       
        \fi}                                    %

\begin{document}

\title{Polarized gluon studies with charmonium and bottomonium at LHCb and AFTER}

\author{Dani\"el Boer}
\email{D.Boer@rug.nl}
\affiliation{Theory Group, KVI, University of Groningen,
Zernikelaan 25, NL-9747 AA Groningen, The Netherlands}

\author{Cristian Pisano}
\email{cristian.pisano@ca.infn.it}
\affiliation{Istituto Nazionale di Fisica Nucleare, Sezione di Cagliari, C.P.\ 170, I-09042 Monserrato (CA), Italy}
\begin{abstract}
Recently it has been put forward that linearly polarized gluons inside unpolarized protons affect the transverse momentum distribution of final state particles in hadronic collisions. They lead to a characteristic modulation of the differential cross section in Higgs production and to azimuthal asymmetries in, for instance, heavy quark pair production. Here we study the effect on charmonium and bottomonium production in hadronic collisions, such as at LHCb and at the proposed fixed target experiment AFTER at LHC. We focus mainly on the scalar and pseudoscalar quarkonia, $\eta_c, \chi_{c0}, \eta_b, \chi_{b0}$, which allow for an angular independent investigation. Within the framework of transverse momentum dependent factorization in combination with the nonrelativistic QCD based color-singlet quarkonium model, we show for small transverse momentum ($q_T^2 \ll 4 M_{Q}^2$) that the scalar and pseudoscalar quarkonium production cross sections are modified in different ways by linearly polarized gluons, while their effects on the production of higher angular momentum quarkonium states are strongly suppressed. Comparisons to $\chi_{c2}, \chi_{b2}$ production  can help to cancel out uncertainties. Together with the analogous study in Higgs production at LHC, quarkonium production can moreover be used to test the scale dependence of the linearly polarized gluon distribution over a large energy range. 
\end{abstract}

\date{\today}

\maketitle

\section{Introduction}

It has been recently pointed out \cite{Brodsky:2012vg,Lansberg:2012kf} that the study of bound states of heavy quarks (quarkonia) with even charge conjugation, such as $\eta_{c, b}$ and $\chi_{c,b}$, produced in high-energy hadronic collisions, provides a useful tool to access the unpolarized gluon distribution function. These quarkonia are produced by fusion of two gluons at scales of the order of their mass, thus large enough to justify the use of perturbative QCD. At leading order, they originate from a $2\to 1$ partonic reaction, with no additional gluon emission in the final state. For this reason event rates should be high and, in analogy to the Drell-Yan process, they have very simple kinematics, with gluon momentum fractions directly related to the rapidity of the observed quarkonium state. This is in contrast to the production mechanisms of $J/\psi$, which has earlier been suggested to pin down the gluon content of the proton \cite{Martin:1987vw,Glover:1987az}. $C=+$ quarkonia should be more reliable as gluon probes, because one expects neither large QCD corrections, nor the many open questions and theoretical uncertainties that affect the predictions of $J/\psi$ and $\Upsilon$ production rates \cite{Brodsky:2012vg,Lansberg:2012kf}. See Ref.\ \cite{Brambilla:2010cs} for a recent review on this topic. QCD corrections to $C=+$ quarkonium production have been calculated in Refs.\ \cite{Kuhn:1992qw,Petrelli:1997ge}.

Hadrons resulting from $2 \to 1$ scattering processes typically have a small transverse momentum, determined by the transverse momenta of the partons in the initial state. Hence, while being mostly lost down the beam pipe at collider facilities like the Large Hadron Collider (LHC), they could be detected by forward detectors (LHCb) or in fixed target experiments (AFTER@LHC). Here we do not aim to make a study of the optimal quarkonium decay channels for these experiments, but according to Ref.\ \cite{Barsuk:2012ic}, the forward detectors and powerful particle identification at LHCb allow for a complete study of all charmonium states through their $p\bar p$ decays. $C$-even charmonium decaying to $\phi\phi$ provides another channel suitable for LHCb, because of the large branching ratio of $\phi\to K^+ K^-$ and the clean signature of two narrow $\phi$ signals \cite{Barsuk:2012ic}. Furthermore, at AFTER, a proposed fixed-target experiment which would utilize the 7-TeV proton beam of the LHC extracted by a bent crystal, yielding collisions at $\sqrt{s} \simeq 115$ GeV, one could study the $\eta_{c,b}$ states in the $\gamma\gamma$ channel and the $\chi_{c,b\,2}$ states through $\ell^+\ell^-\gamma$ decays \cite{Brodsky:2012vg,Lansberg:2012kf}. For discussions of promising channels to study $\eta_b$ production, cf.\ e.g.\ Refs.\ \cite{Maltoni:2004hv,Braguta:2005gw,Hao:2006nf}.

The states that we will mainly focus on in this paper will be the even charge conjugation, scalar and pseudoscalar quarkonia, i.e.\ $J^{PC}=0^{\pm +}$ states, or in terms of spectroscopic notation ($^{2S+1}L_J$) the $^{1}S_0$ states $\eta_c, \eta_b$ and $ ^{3}P_0$ states $\chi_{c0}, \chi_{b0}$.  
The $\chi_{c1}, \chi_{b1}$ states would require a different treatment, because they suffer from the same problem as other vector states, such as the $J/\psi$, namely that because of the Landau-Yang theorem, their production from two gluons requires either an additional gluon radiated off or inclusion of the off shellness of gluons. Although off-shell gluons contribute significantly to $\chi_{c1}$ production according to Ref.\ \cite{Hagler:2000dd}, the theoretical treatment is more involved. Moreover, the contribution of linearly polarized gluons is suppressed in $\chi_{c1}$ or $\chi_{b1}$ production due to helicity conservation. An even stronger suppression arises for $\chi_{c2}$ and $\chi_{b2}$ states, but as these {\em can} be produced from two on-shell gluons, they are theoretically cleaner and can help to cancel out uncertainties as we will discuss.

Recently it was pointed out \cite{Boer:2011kf,Dunnen:2012ym} that in the production of Higgs bosons at LHC, there is in principle also a contribution from linearly polarized gluons inside unpolarized hadrons \cite{Mulders:2000sh}. This affects the transverse momentum distribution in such a characteristic way that it provides a tool to determine whether the Higgs is a scalar or pseudoscalar boson. As the distribution of linearly polarized gluons inside unpolarized protons is currently unknown (only a theoretical upper bound has been derived \cite{Mulders:2000sh,Boer:2010zf}), no model independent predictions for the modification of the transverse momentum distribution could be given. For this purpose an independent experimental determination is required and several suggestions have been put forward, such as through azimuthal asymmetries in heavy quark pair production in deep inelastic electron-proton scattering (with the heavy quarks having large relative and absolute transverse momenta) \cite{Boer:2010zf} and photon pair production in proton-proton collisions \cite{Qiu:2011ai}. Here we will discuss another promising method to access and extract the distribution of linearly polarized gluons inside unpolarized hadrons, namely that of (pseudo-)scalar charmonium or bottomonium production, which is more closely related to Higgs production and has the additional advantage of providing a way to map out the scale dependence. The latter is dictated by the framework of transverse momentum dependent (TMD) factorization, which for Higgs production including linearly polarized gluons has been discussed in Ref.\ \cite{Sun:2011iw}. In a collinear factorization approach, polarized gluons are generated perturbatively \cite{Balazs:2007hr,Nadolsky:2007ba,Catani:2010pd}, while in the framework of TMD factorization polarized gluons are already present at tree level through a nonperturbative distribution, here denoted by $h_1^{\perp\, g}$. It corresponds to an interference between $+1$ and $-1$ helicity gluon states that would be suppressed without transverse momentum. As a consequence, $h_1^{\perp\, g}$ shows up in transverse momentum distributions, either of pairs of particles or of single particles. It can modulate the angular part of the transverse momentum distribution, but  
it can also generate a term in the cross section that is independent of the azimuthal angle. This happens when two linearly polarized  gluons, one from each hadron, participate in the scattering and the final state particle is a scalar or pseudoscalar. In this paper we will study how the distribution of linearly polarized gluons may affect the angular-independent transverse momentum distributions of $0^{\pm +}$ quarkonia, employing TMD factorization in combination with the nonrelativistic color-singlet quarkonium model, which we will justify
from the framework of nonrelativistic QCD (NRQCD). NRQCD in combination with TMD factorization, has been used in Ref.\ \cite{Yuan:2008vn} to study single spin asymmetries in quarkonium production in lepton-nucleon and nucleon-nucleon collisions. It was observed that the asymmetry is nonzero in $ep^\uparrow$ collisions only in the color-octet model, whereas in $pp^\uparrow$ collisions only in the color-singlet model. In Ref.\ \cite{Godbole:2012bx} the asymmetry for $J/\psi$ production in $ep^\uparrow$ collisions was studied using the color evaporation model.

In the color-singlet model (CSM) \cite{Kuhn:1979bb,Guberina:1980dc,Baier:1983va}, the quarkonium bound state is produced in a color-singlet state from a heavy quark and antiquark with small relative momenta. Although the CSM fails to describe the large transverse momentum spectra of $C=-$ vector states $J/\psi$, $\psi(2S)$, and $\Upsilon$ (for a discussion cf.\ Refs.\ \cite{Bodwin:2005hm,Brambilla:2010cs}), it can be justified from NRQCD for $C=+$ states, like for $\eta_{b}$ in 
Ref.\ \cite{Maltoni:2004hv}.  
In the framework of NRQCD \cite{Bodwin:1994jh,Cho:1995vh,Cho:1995ce}, hadroproduction of quarkonia is described in terms of a double power series expansion in the strong coupling constant $\alpha_s$  and in the velocity parameter $v$, such that $(M_Qv^2)^2 \ll (M_Qv)^2 \ll M_Q^2$, with $M_Q$ the heavy quark mass. The magnitude of the velocity is given by the self-consistency condition $v \sim \alpha_s(M_Qv)$, yielding $v^2 \simeq 0.3$ for charmonium and $v^2 \simeq 0.1$ for bottomonium. For details we refer to Refs.\ \cite{Bodwin:1994jh,Cho:1995vh,Cho:1995ce}.
According to NRQCD, a heavy quark-antiquark pair can be produced at short distances not just as a color-singlet, but also in a color-octet configuration, which subsequently evolves into a physical quarkonium state by radiating soft gluons. The hadronization of the pair is encoded in long-distance, universal, matrix elements that can be characterized according to the velocity expansion in $v$. These matrix elements are not calculable perturbatively and have to be determined by a fit to the data. Lattice QCD estimates are available only for annihilation matrix elements \cite{Bodwin:2005gg}. NRQCD has had success in explaining many experimental observations, but it is not able to reproduce in a consistent way all cross sections and polarization measurements for charmonia, see Refs.\ \cite{Brodsky:2012vg,Brambilla:2010cs,Lansberg:2012kf} and references therein. Nevertheless, for the low transverse momentum part of the spectrum of $C=+$ states, we expect it to provide a reasonable description, such as for the $\chi_{c1,2}$ states discussed in Ref.\ \cite{Bodwin:2005hm}. For $C=+$ states an NRQCD analysis \cite{Bodwin:1994jh} shows that color-octet contributions are (at least) order $v^2$ suppressed with respect to the singlet contributions. Earlier, in a study \cite{Hagler:2000dd} of hadroproduction of $\chi_{cJ}$ in the $k_\perp$-factorization approach, it was also noted that the color-octet part is strongly suppressed and the color-singlet part dominates, especially at low transverse momenta. Also, according to Refs.\ \cite{Brodsky:2012vg,Lansberg:2012kf}, color-octet contributions can certainly be neglected for $C=+$ bottomonium, in agreement with the analysis of Ref.\ \cite{Maltoni:2004hv}. At large transverse momentum \cite{Mathews:1998nk,Biswal:2010xk} and central exclusive production \cite{Pasechnik:2007hm} complications may arise, but these are beyond the scope of this paper. At low transverse momenta we consider it justified to restrict to the color-singlet contributions for our particular analysis, which is primarily aimed at pointing out how the linear polarization of gluons enters in the production cross sections of $C=+$ quarkonium states. 

\section{Outline of the calculations}

\begin{figure}[t]
\begin{center}
\epsfig{file=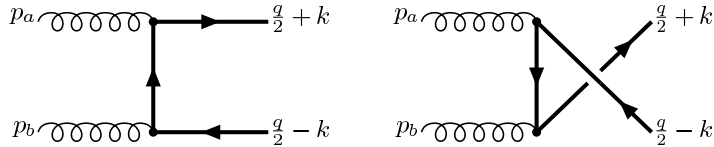}
\end{center}
\caption{\it Feynman diagrams for the leading order gluon-gluon fusion process in the color-singlet model.} 
\label{fig:gglo}
\end{figure}
We consider the process 
\begin{equation}
h(P_A){+}h(P_B)\to Q \bar{Q} [^{2S+1}L^{(1)}_J ] (q){+}X \, ,
\label{eq:proc}
\end{equation}
where the four-momenta of the particles are given within round brackets, and we assume that 
the colorless heavy quark-antiquark pair is in a bound state described by a nonrelativistic wave function 
with spin $S$, orbital angular momentum $L$ and total angular momentum $J$. The $S$, $L$, and $J$ quantum numbers are indicated within the square brackets in spectroscopic notation, while the color assignments of the pair are generally specified by singlet or octet superscripts, $(1)$ or $(8)$. 
The squared invariant mass of the $Q \bar{Q}$ state is $M^2=q^2$ and $M = 2\,M_Q$ up to small relativistic corrections.  According to the CSM \cite{Kuhn:1979bb,Guberina:1980dc,Baier:1983va}, we assume that the two quarks are produced in a hard partonic scattering with the same quantum numbers as the mesons into which they nonperturbatively evolve. Therefore, to lowest order in perturbative QCD, ${{\cal{O}}(\alpha_s^2)}$, one only has the gluon fusion process described by the two Feynman diagrams depicted in Fig.\ \ref{fig:gglo}.
This production mechanism is only significant in the low $q_\sT$ region with a mean $q_\sT$ determined by the intrinsic transverse motion of the gluon constituents. In practice, a fixed order calculation is not expected to describe very accurately the small transverse momentum spectrum \cite{Maltoni:2004hv}, but corrections and resummation can be considered at a later stage when it becomes relevant. Moreover, we expect that in the {\it comparison} of different quarkonia the effect of higher order corrections will matter less. 
Another aspect to keep in mind is that at small transverse momentum, but large $x_F$, higher Fock states are expected to become relevant, such as considered in forward Higgs production in Ref.\ \cite{Brodsky:2007yz}.  

Transverse momentum dependent gluon distributions inside 
unpolarized hadrons are defined via a matrix element of a correlator of the gluon field strengths $F^{\mu \nu}(0)$ and
$F^{\nu \sigma}(\xi)$, evaluated at fixed light-front (LF) time $ \xi^+ = \xi{\cdot}n=0$, with $n$ being a lightlike vector conjugate to the four-momentum $P$ of the parent hadron. Expanding the gluon four-momentum as 
$p = x\,P + p_\sT + p^- n$, the correlator can be written as \cite{Mulders:2000sh}
\begin{eqnarray}
\label{GluonCorr}
\Phi_g^{\mu\nu}(x,\bm p_\sT )
& = &  \frac{n_\rho\,n_\sigma}{(p{\cdot}n)^2}
{\int}\frac{d(\xi{\cdot}P)\,d^2\xi_\sT}{(2\pi)^3}\
e^{ip\cdot\xi}\,
\langle P|\,\tr\big[\,F^{\mu\rho}(0)\,
F^{\nu\sigma}(\xi)\,\big]
\,|P \rangle\,\big\rfloor_{\text{LF}} \nonumber \\
& = &
-\frac{1}{2x}\,\bigg \{g_\sT^{\mu\nu}\,f_1^g (x,\bm p_\sT^2)
-\bigg(\frac{p_\sT^\mu p_\sT^\nu}{M_h^2}\,
{+}\,g_\sT^{\mu\nu}\frac{\bm p_\sT^2}{2M_h^2}\bigg)
\;h_1^{\perp\,g} (x,\bm p_\sT^2) \bigg \} ,
\end{eqnarray}
where $p_{\sT}^2 = -\bm p_{\sT}^2$, $g^{\mu\nu}_{\sT} = g^{\mu\nu}
- P^{\mu}n^{\nu}/P{\cdot}n-n^{\mu}P^{\nu}/P{\cdot}n$ and $M_h$ is the hadron mass. The unpolarized and linearly polarized gluon distributions are denoted by 
$f_1^g(x,\bm{p}_\sT^2)$ and 
$h_1^{\perp\,g}(x,\bm{p}_\sT^2)$, respectively. A Wilson line, which makes the correlator gauge invariant, has been omitted  in Eq.\ (\ref{GluonCorr}). Like other TMDs, also $h_1^{\perp\, g}(x, \bm p_\sT)$ receives 
contributions from initial and/or final state interactions that render the gauge link process dependent. 
Therefore, in principle $h_1^{\perp\, g}(x, \bm p_\sT^2)$ can 
receive nonuniversal contributions, despite the fact that it is $T$ even, and its extraction can be hampered in those processes for which factorization does 
not hold, {\it e.g.} dijet production in hadron-hadron collisions \cite{Collins:2007nk,Boer:2009nc,Rogers:2010dm,Buffing:2011mj}. For the present process we expect no problems with factorization since the $C$-even quarkonium bound state forms a color singlet. In this case there is only one color trace present and gauge links can always be factorized.  

Along the lines of Refs.\ \cite{Boer:2007nd,Boer:2009nc,Boer:2011kf}, we assume that at sufficiently high energies TMD factorization holds and the cross section for the process in Eq.\ (\ref{eq:proc}) is given by
\begin{eqnarray}
\d\sigma
& = &\frac{1}{2 s}\,\frac{d^3 \bm q}{(2\pi)^3\,2 q^0} 
{\int} \d x_a \,\d x_b \,\d^2\bm p_{a\sT} \,\d^2\bm p_{b\sT}\,(2\pi)^4
\delta^4(p_a{+} p_b {-} q)
 \nonumber \\
&&\qquad \qquad\qquad \qquad\qquad\times
{\rm Tr}\, \left \{ \Phi_g(x_a {,}\bm p_{a \sT}) \Phi_g(x_b {,}\bm p_{b \sT})\overline{\sum_{\rm colors}}
 \left|{\cal A} \left (g\, g  \rightarrow Q \bar{Q} [^{2S+1}L^{(1)}_J ] \right ) (p_a, p_b; q)\right|^2\right \}\,,
\label{CrossSec}
\end{eqnarray}
where $s = (P_A + P_B)^2$ is the total energy squared in the hadronic 
center-of-mass frame. Here and in following expressions, $p_a$ and $p_b$ are understood to have $p_a^-=p_b^+ =0$. The amplitude ${\cal A}$ can be written in the form \cite{Kuhn:1979bb,Guberina:1980dc,Baier:1983va}
\begin{equation}
{\cal{A}} \left (g\, g  \rightarrow Q \bar{Q} [^{2S+1}L^{(1)}_J ] \right ) (p_a, p_b; q) = \int \frac{\d^4k}{(2\pi)^4}\, {\rm Tr}\left [O(p_a,p_b; k)\,\phi_{L S; J J_z} (q,k)\right]\,,
\label{eq:ampl}
\end{equation}
with $2 k$ being the relative momentum of the heavy quarks, see Fig.\ \ref{fig:gglo}. In Eq.\ (\ref{eq:ampl}) $\phi_{L S; J J_z} (q,k)$ is the Bethe-Salpeter wave function of the produced bound state and $O(p_a,p_b; k)$ is calculated from the Feynman diagrams in Fig.\ \ref{fig:gglo} without including the polarization vectors of the initial gluons and the heavy quark legs, since they are absorbed into the definitions of the gluon correlators and of the bound state wave function, respectively. Therefore we have
\begin{eqnarray}
O^{\mu\nu}(p_a, p_b; k) & = & 2 g_s^2\, \sum_{i,j}\,\langle 3i;\bar{3}j\vert 1\rangle \,\left \{ (T^a T^b)_{ij} \,\gamma^\mu\,\frac{\ps_b-\ps_a + 2 \ks + 
2 M_Q}{(p_a-p_b-2k)^2-4 M_Q^2}\,\gamma^\nu \right .\nonumber \\
& & \hspace{4.cm}+ \,\left . (T^b T^a)_{ij} \,\gamma^\nu\,\frac{\ps_a-\ps_b + 2 \ks + 
2 M_Q}{(p_a-p_b+2k)^2-4 M_Q^2}\,\gamma^\mu \right \}\,,
\label{eq:O}
\end{eqnarray}
where the sum is taken over the colors of the outgoing quark and antiquark and $T^a$ are the $SU(3)$ matrices normalized according to Tr$(T^aT^b)=$ Tr$(T^bT^a) = \delta^{ab}/2$. The $SU(3)$ Clebsch-Gordan coefficients,
\begin{equation}
\langle 3i;\bar{3}j\vert 1\rangle = \frac{\delta^{ij}}{\sqrt{N_c}}\,,
\end{equation}
with $N_c$ being the number of colors, project out the color-singlet configuration. In the rest frame of the bound state, the relative momentum of the two quarks is small compared to their mass $M_Q$, which justifies a nonrelativistic approach. Hence $\phi_{L S; J J_z}$ can be written as
\begin{equation}
\phi_{L S; J J_z} (q,k) = 2 \pi \delta \left (k^0-\frac{\bm k^2}{ M} \right )\sum_{L_z, S_z} \Psi_{L L_z}(\bm k) \langle L L_z; S S_z \vert J J_z \rangle {\cal P}_{S S_z} (q,k)\,,
\label{eq:wf}
\end{equation}
where $\Psi_{L L_z}(\bm k)$ is the eigenfunction of the orbital angular momentum $L$ and the brackets denote the appropriate Clebsch-Gordan coefficients. The spin projection operator ${\cal P}_{S S_z}$ is defined in terms of the heavy quark and antiquark spinors, $u\left (q/2 + k, s \right )$ and 
$\bar {v} \left ( q/2-k , \bar s \right ) $, respectively,  with $s$ and $\bar s$ being their spins.
More explicitly,
\begin{eqnarray}
{\cal P}_{S S_z} (q,k) & \equiv & \sum_{s,\bar s} \langle \frac{1}{2}\, s; \frac{1}{2}\,\bar s\vert S S_z \rangle u \left (\frac{q}{2} + k, s \right ) \bar v \left (\frac{q}{2}-k , \bar s \right )  \nonumber \\
& = & \frac{1}{4 M^{3/2}}\, \left ( \qs   + 2 \ks + M \right ) \Pi_{SS_z}\left ( -{\qs }  + 2 \ks + M \right ) + {\cal O}(\bm k^2)\,,
\label{eq:spinpr}
\end{eqnarray}
with  $\Pi_{SS_z} = \gamma^5$ for singlet ($S=0$) states and $\Pi_{SS_z} = \epss_{S_z} (q)$ for triplet  ($S=1$)  states, where $\varepsilon_{S_z}(q)$ is the spin polarization vector of the $Q\bar Q$ system.
Substituting Eq.\ (\ref{eq:wf}) into Eq.\ (\ref{eq:ampl}), one gets 
\begin{equation}
{\cal{A}} \left (g\, g  \rightarrow Q \bar{Q} [^{2S+1}L^{(1)}_J ] \right ) (p_a, p_b; q) = \sum_{L_z, S_z}\,\int \frac{\d^3 \bm k}{(2\pi)^3}\, \Psi_{L L_z}(\bm k) \langle L L_z; S S_z \vert J J_z \rangle {\rm Tr}\left [O(p_a,p_b; k)\,\phi_{L S; J J_z} (q,k)\right]\,,
\label{eq:ampl2}
\end{equation}
which can be expanded in powers of $\vert \bm k \vert$ around $\bm k =0$. At this point it is 
convenient to separate the Fourier transform of the wave function $\Psi_{LL_z} (\bm k)$
into its radial and angular pieces,
\begin{equation}
\int \frac{\d^3 \bm k}{(2\pi)^3}\, e^{i \bm k \cdot \bm r}\,\Psi_{L L_z}(\bm k) = \tilde \Psi_{L L_z}(\bm r) = R_L(\vert \bm r\vert )\, Y_{LL_z}(\theta,\varphi)\,,
\label{eq:ft}
\end{equation}
where $\bm r = (\vert \bm r\vert,\theta,\varphi)$ in spherical coordinates, $R_L (\vert \bm r\vert)$ is the 
radial wave function and $Y_{LL_z}(\theta,\varphi)$  is a spherical harmonic. This implies, in particular,
\begin{equation}
\int\frac{\d^3 \bm k}{(2\pi)^3}\,\Psi_{00}(\bm k) = \frac{1}{\sqrt{4\pi}}\,R_0(0)\,.
\label{eq:R00}
\end{equation}
Therefore, for $S$ waves ($L=0$, $J=0,1$), the first term of the Taylor expansion of the amplitude is obtained by calculating the integrand in Eq.\ (\ref{eq:ampl2}) at $\bm k = 0$ and, 
since this does not depend on $\bm k$ anymore, it can be factored out and one is left with the integral in Eq.\  (\ref{eq:R00}). Making use of Eq.\ (\ref{eq:spinpr}) as well,
one gets 
\begin{eqnarray}
{\cal{A}} [^{2J+1}S^{(1)}_{J} ] (p_a, p_b; q) = \frac{1}{\sqrt{4 \pi}}\, R_0(0)\, {\rm Tr}\left [O(p_a,p_b; 0)\, {\cal P}_{S S_z} (q,0) \right ]\,,
\end{eqnarray}
which, adopting the definition
${O}(0) \equiv O(p_a,p_b; 0)$ with $p_b = q-p_a$, leads to the final results \cite{Baier:1983va}:
\begin{eqnarray}
{\cal{A}} [^{1}S^{(1)}_{0} ] (p_a, q) &  = &   \frac{1}{4 \sqrt{\pi\, M }}\, R_0(0)\, {\rm Tr}\left [O(0)\, (\qs + M) \,\gamma^5 \right ]\,,\label{eq:1S0} \\
{\cal{A}} [^{3}S^{(1)}_{1} ] (p_a, q) &  = &   \frac{1}{4 \sqrt{\pi\, M }}\, R_0(0)\, {\rm Tr}\left [O(0)\, (\qs + M) \,\epss_{S_z} \right ]\,.\label{eq:3S1}
\end{eqnarray} 
For $P$ waves ($L=1$, $J=0, 1, 2$) $R_1(0) =0$, so one needs to consider the linear term in $k^\alpha$ in the expansion of Eq.\ (\ref{eq:ampl2}). Again from Eq.\ (\ref{eq:ft}) it can be shown that 
\begin{equation}
\int\frac{\d^3 \bm k}{(2\pi)^3}\,k^\alpha \,\Psi_{1 L_z}(\bm k) = -i\, \varepsilon^{\alpha}_{L_z}(q)\,\sqrt{\frac{3}{4\pi}}\,
\,R_1^\prime (0) \,,
\end{equation}
 where $\varepsilon_{L_z}^{\alpha}(q)$ is a polarization vector referring to an $L=1$ bound state, and $R_1^\prime(0)$ is the derivative of the $P$-wave 
(radial) wave function evaluated at the origin. One gets \cite{Baier:1983va}
\begin{eqnarray}
{\cal{A}}[^{2S+1}P^{(1)}_{J} ](p_a, q) &  = & -i \,\sqrt{\frac{3}{4\pi}}\, R_1^{\prime}(0)\, \sum_{L_z, S_z}\, 
\langle 1 L_z ; S S_z\vert J J_z\rangle \, \varepsilon^\alpha_{L_z}(q)\,\left. \frac{\partial }{\partial k^{\alpha}}
\, {\rm Tr} \left [O(p_a, q;  k)\, 
{\cal P}_{S S_z} (q,k) \right ] \right \vert_{ k=0}.
\label{eq:P}
\end{eqnarray}
The numerical values of the radial wave functions and their derivatives at the origin  are commonly obtained from potential models, or from leptonic decay matrix elements \cite{Eichten:1995ch,Bodwin:2007fz}. For the singlet state, Eq.\ (\ref{eq:P}) reduces to
\begin{equation}
{\cal{A}} [^{1}P^{(1)}_{0} ] (p_a, q)  =  -i \sqrt{\frac{3 }{4\pi M}} \, R_1^\prime(0)\, 
{\rm Tr}\left [ \left ({O}(0)\,\epss_{L_z}\frac{\qs}{M} +   \varepsilon^{\alpha}_{L_z} \hat{O}_{\alpha}(0)\, \frac{\qs+M}{2} \right ) \gamma^5 \right ]\,,\label{eq:1P0}
\end{equation}
where we have defined
\begin{equation}
 \hat {O}_\alpha(0) \equiv \, \left . \frac{\partial }{\partial k^{\alpha}}{O}(p_a, q; k)\right  \vert_{ k=0}.
\end{equation}
In order to write the triplet amplitudes in an analogous form, we utilize the following relations for the Clebsch-Gordan coefficients and the various polarization vectors \cite{Kuhn:1979bb,Guberina:1980dc}:
\begin{eqnarray}
\sum_{L_z, S_z} \,\langle 1 L_z ; 1 S_z\vert 0 0\rangle \, \varepsilon^\alpha_{S_z}(q) \,\varepsilon^\beta_{L_z}(q) & = & \sqrt{\frac{1}{3}}\,\left (g^{\alpha \beta} -\frac{q^\alpha q^\beta}{M^2}\right )\, , \nonumber \\
\sum_{L_z, S_z}\, \langle 1 L_z ; 1 S_z\vert 1 J_z\rangle \, \varepsilon^\alpha_{S_z}(q) \, \varepsilon^\beta_{L_z}(q) & = & -\frac{i}{M}\,\sqrt{\frac{1}{2}}\,{\epsilon_{\mu\nu\rho\sigma}}\,g^{\rho\alpha}\,g^{\sigma\beta}\,q^{\mu}\,\varepsilon^\nu_{J_z}(q)\,, \nonumber \\
\sum_{L_z, S_z}\, \langle 1 L_z ; 1 S_z\vert 2 J_z\rangle \, \varepsilon^\alpha_{S_z}(q) \, \varepsilon^\beta_{L_z}(q) & = & \varepsilon^{\alpha\beta}_{J_z}(q)\,.
\end{eqnarray}
Hence, from Eq.\ (\ref{eq:P}), it is straightforward to express the triplet amplitudes as follows \cite{Kuhn:1979bb,Guberina:1980dc}:
\begin{eqnarray}
{\cal{A}} [^{3}P^{(1)}_{0} ] (p_a, q) &  = & -i \sqrt{\frac{1}{4\pi M}}\,R_1^\prime(0)\,{\rm Tr}\left [ 3\,O(0) + \left (\gamma_\alpha \hat{O}^\alpha(0) -\frac{\qs \,q_\alpha}{M^2}\, \hat{O}^\alpha(0) \right ) \frac{\qs + M}{2} \right ]\,, \label{eq:3P0} \\ 
{\cal{A}} [^{3}P^{(1)}_{1}] (p_a, q) &  = & - \sqrt{\frac{3}{8\pi M}} \, R_1^\prime(0)\,  
\epsilon_{\mu\nu\alpha\beta}\,\frac{q^\mu}{M}\,\varepsilon_{J_z}^\nu(q)\, {\rm Tr} \left [\gamma^\alpha \hat{O}^\beta(0) \,\frac{\qs + M}{2}   + O(0)\, \frac{\qs}{M}\,\gamma^\alpha\gamma^\beta \right ]
\label{eq:3P1}   \,,\\
{\cal{A}} [^{3}P^{(1)}_{2}] (p_a, q) & = & -i\sqrt{\frac{3}{4\pi M}}\, R_1^\prime(0)\, 
\varepsilon^{\alpha\beta}_{J_z}(q)\, {\rm Tr}\left [ \gamma_\alpha\,{\hat O}_\beta(0)\,
\frac{\qs + M}{2} \right ]\,.
\label{eq:3P2}
\end{eqnarray}
Here $\varepsilon^{\nu}_{J_z}$ is the polarization vector for a bound state with $J=1$ and it obeys the usual relations (which hold also for $\varepsilon^\nu_{S_z}$ and  $\varepsilon^\nu_{L_z}$),
\begin{equation}
\varepsilon_{J_z}^\alpha(q) \,q_\alpha = 0\,, \qquad  \sum_{S_z} \varepsilon^\alpha_{J_z}(q)\, \varepsilon^{*\beta}_{J_z}(q) = -g^{\alpha\beta}+\frac{q^\alpha q^\beta}{M^2} \equiv {\cal Q}^{\alpha\beta}\,,
\label{eq:eps}
\end{equation}
while $\varepsilon^{\alpha\beta}_{J_z}$ is the polarization tensor for a spin 2 system and satisfies
\begin{eqnarray}
&&\qquad \varepsilon_{J_z}^{\alpha\beta}(q) = \varepsilon_{J_z}^{\beta\alpha}(q) \,, 
\qquad {\varepsilon_{J_z}^{\alpha}}_\alpha (q) = 0\, , \qquad  q_\alpha \varepsilon_{J_z}^{\alpha\beta}(q) =0\,, \nonumber \\ 
&&\varepsilon_{J_z}^{\mu\nu}(q)  \,\varepsilon_{J_z}^{*\alpha\beta}(q)   =  
\frac{1}{2}\, \left [ {\cal Q}^{\mu\alpha} {\cal Q}^{\nu\beta} + {\cal Q}^{\mu\beta}{\cal Q}^{\nu\alpha} \right ] - \frac{1}{3}\, {\cal Q}^{\mu\nu} {\cal Q}^{\alpha\beta}\,.
\label{eq:eps2}
\end{eqnarray}
By substituting the explicit expressions of the operators $O(0)$ and $\hat{O}^\alpha(0)$, 
\begin{eqnarray}
O^{\mu\nu}(0) & = & -\frac{\delta^{ab}}{\sqrt{N_c}}\, \frac{g_s^2}{2M^2}\, \left [ \gamma^\mu \left ( \qs -2\,\ps_a + M \right )\gamma^\nu \  - \gamma^\nu \left ( \qs -2\,\ps_a - M \right )\gamma^\mu \right ]  \, ,\nonumber \\
\hat{O}^{\alpha\mu\nu}(0) & = & \frac{\delta^{ab}}{\sqrt{N_c}}\, \frac{g_s^2}{M^2}\, \left \{  \frac{2 p_a^ \alpha}{M^2}   \left [ \gamma^\mu \left ( \qs -2\,\ps_a + M \right )\gamma^\nu \  + \gamma^\nu \left ( \qs -2\,\ps_a - M \right )\gamma^\mu \right ] - \gamma^\mu\gamma^\alpha\gamma^\nu - \gamma^\nu\gamma^\alpha\gamma^\mu \right \}\,, 
\end{eqnarray}
derived from Eq.\ (\ref{eq:O}) into Eqs.\ (\ref{eq:1S0}), (\ref{eq:3S1}), (\ref{eq:1P0}) and (\ref{eq:3P0})-(\ref{eq:3P2}), it is  found that the only nonzero amplitudes are the ones corresponding to the $^1S_0$ ($\eta_Q$) and  $^3P_{0,2}$ ($\chi_{Q0,2}$) states, where $Q = c, b$:
\begin{eqnarray}
{\cal{A}}^{\mu\nu} [^{1}S^{(1)}_{0} ] (p_a, q) &  = & -2i\, \frac{\delta^{a b}}{\sqrt{N_c}}\,  \frac{g_s^2}{\sqrt{\pi M^5}}\,R_0(0)\, \epsilon^{\mu\nu\rho\sigma}p_{a\rho} q_{\sigma}\,, \\
{\cal{A}}^{\mu\nu} [^{3}P^{(1)}_{0} ] (p_a, q) &  = & -2i\, \frac{\delta^{a b}}{\sqrt{N_c}}\,  \frac{g_s^2}{\sqrt{\pi M^3}}\,R_0^\prime(0)\, \left [ -3 g^{\mu\nu} + \frac{2}{M^2}\, q^\mu p_a^{\nu}\right ]\, ,\\
{\cal{A}}^{\mu\nu} [^{3}P^{(1)}_{2} ] (p_a, q) &  = & -2i\, \frac{\delta^{a b}}{\sqrt{N_c}}\,   \sqrt{\frac{3}{\pi M^3}}\, g_s^2\,R_0^\prime(0)\,\varepsilon_{J_z}^{\rho\sigma}(q)
\left [ \frac{4}{M^2}\,g^{\mu\nu} p_{a\rho} p_{a\sigma}  - g^{\mu}_\rho g^{\nu}_\sigma - g^{\nu}_\rho g^{\mu}_\sigma \right ]\,.
\end{eqnarray}
After squaring and averaging over the colors of the incoming gluons, the corresponding cross sections are calculated using Eq.\ (\ref{CrossSec}). 

Here the cross sections will be written in terms of operator matrix elements appearing in NRQCD. The long-distance matrix element of relevance for the $\eta_Q$ is related to the $\eta_Q$ radial wave function at the origin by the following relation 
\cite{Bodwin:1994jh}:
\begin{equation}
\langle 0\vert {\cal{O}}^{\eta_Q}_1 (^1 S_0)\vert 0\rangle = \frac{N_c}{2\pi}\, \vert R_0(0)\vert^2 [1+{\cal{O}}(v^4)]\,,
\label{meeta}
\end{equation}
where $v$ is the velocity of the heavy quarks inside the $Q\bar Q$ bound 
state. The other color singlet matrix element, $\langle 0\vert {\cal{P}}^{\eta_Q}_1 (^1 S_0)\vert 0\rangle$, 
starts out order $v^2$ suppressed and will be dropped. The color-octet contributions $^1 S_0^{(8)}$, $^3 S_1^{(8)}$, $^1 P_1^{(8)}$ are, according to the NRQCD counting rules, suppressed by $v^4, v^3$, and $v^4$, respectively \cite{Bodwin:1994jh,Bodwin:2005hm} with respect to the leading color-singlet term and will also not be included. 

Similarly, for the $\chi_{QJ}$ states the matrix element of relevance is \cite{Bodwin:1994jh}
\begin{equation}
\langle 0\vert {\cal{O}}^{\chi_{Q J}}_1 (^3 P_J)\vert 0\rangle = \frac{3 N_c}{2\pi}\, \vert R_1^{\prime}(0)\vert^2 [1+{\cal{O}}(v^2)]\, , \quad J = 0,1,2\,.
\label{mechi}
\end{equation}
This matrix element is order $v^2$ suppressed with respect to the matrix element in Eq.\ (\ref{meeta}) for $\eta_Q$ \cite{Cho:1995vh}, 
but is the leading one for $\chi_{QJ}$ states. However, according to the NRQCD counting rules, there is also a color-octet contribution of the same order in $v^2$. In Ref.\ \cite{Bodwin:2005hm} it is shown that its matrix element is nevertheless suppressed, as reflected by the ratio $R^{\chi_Q} \equiv M_Q^2
\langle {\cal{O}}^{\chi_{QJ}}_8 (^3 S_1)\rangle/ \langle {\cal{O}}^{\chi_{QJ}}_1 (^3 P_J)\rangle \approx v^0/(2N_c) \approx 0.17$, where now the suppression comes from the color factor $N_c$. The suppression of the color-octet contribution in $\chi_c$ production 
was also noted in Ref.\ \cite{Hagler:2000dd}. 

Expressing the CSM results in terms of the leading color-singlet NRQCD matrix elements in Eqs.\ (\ref{meeta}) and (\ref{mechi}), yields the production cross sections presented in the next section for $J=0$ and $J=2$ states. For $J=1$ there is no contribution. 

\section{Quarkonium production cross sections at small transverse momentum}    

The calculation as explained in the previous section yields our main results for the production cross sections of $J=0$ and $J=2$ $C=+$ quarkonia:
 \begin{eqnarray}
\frac{\d\sigma(\eta_Q)}{\d y\,\d^{2}\bm{q}_\sT} & = & \frac{2}{9} \frac{\pi^3\alpha_s^2}{M^3\,s}\, \langle 0\vert {\cal{O}}^{\eta_Q}_1 (^1 S_0)\vert 0\rangle\,  \mathcal{C}\left[f_{1}^{g}\, f_{1}^{g}\right]\,\left [1 - R(\bm q_\sT^2) \right]\,,\label{eq:CSeta}
\end{eqnarray} 
\begin{eqnarray}
\frac{\d\sigma (\chi_{Q 0})}{\d y\,\d^{2}\bm{q}_\sT} & = &   \frac{8}{3} \frac{\pi^3\alpha_s^2}{ M^5\,s}\, \langle 0\vert {\cal{O}}^{\chi_{Q0}}_1 (^3 P_0)\vert 0\rangle\,\mathcal{C}\left[f_{1}^{g}\, f_{1}^{g}\right]\, \left [ 1 + R(\bm q_\sT^2)\right]\, ,\label{eq:CSchi0}
 \end{eqnarray}
\begin{eqnarray}
\frac{\d\sigma (\chi_{Q 2})}{\d y\,\d^{2}\bm{q}_\sT} & = &   \frac{32}{9} \frac{\pi^3\alpha_s^2}{ M^5\,s}\, \langle 0\vert {\cal{O}}^{\chi_{Q2}}_1 (^3 P_2)\vert 0\rangle \,\mathcal{C}\left[f_{1}^{g}\, f_{1}^{g}\right] \, ,\label{eq:CSchi2}
\end{eqnarray}
with $N_c=3$ and corrections of ${\cal{O}}(\bm q_\sT^2/M^2)$ and ${\cal{O}}(v^2)$.
Here $y$ is the rapidity of the produced bound state along the direction of the incoming hadrons and $R(\bm q_\sT^2)$ is defined as the ratio
\begin{equation}
R (\bm q_{\sT}^2) \equiv \frac{\mathcal{C}\left[w\, h_{1}^{\perp \,g}\, h_{1}^{\perp \,g} \right ]}{\mathcal{C}\left[f_{1}^{g}\, f_{1}^{g}\right]}~.
\end{equation}
The TMD convolutions are given by 
\begin{eqnarray}
\mathcal{C}[w\, f\, f] & \equiv & \int d^{2}\bm p_{a\sT}\int d^{2}\bm p_{b\sT}\,
\delta^{2}(\bm p_{a\sT}+\bm p_{b\sT}-\bm q_{\sT})\, w(\bm p_{a\sT},\bm p_{b\sT})\, f(x_{a},\bm p_{a\sT}^{2})\, f(x_{b},\bm p_{b\sT}^{2})\,,\label{eq:Conv}
\end{eqnarray} 
with the transverse momentum weight
\begin{equation}
w= \frac{1}{2M_h^{4}} \, \left [ (\bm p_{a\sT}\cdot\bm p_{b\sT})^{2}-\frac{1}{2} \, \bm p_{a\sT}^{2}
\bm p_{b\sT}^{2} \right ]\label{eq:weight}\,,
\end{equation}
and, neglecting again terms of ${\cal{O}} \left ( {\bm q_\sT^2}/{M^2} \right )$,
\begin{equation}
x_a =  \frac{M}{\sqrt{s}}\, e^y \,,  \qquad x_b = \frac{M}{\sqrt{s}}\,e^{-y} \,.
\end{equation}
The results in Eqs.\ (\ref{eq:CSeta})-(\ref{eq:CSchi2}), integrated over $\bm q_\sT$, are in agreement with the ones published in Ref.\ \cite{Baier:1983va} and obtained without the inclusion of intrinsic transverse momenta of the incoming gluons. Furthermore, we note that $\int d^{2}\bm q_{\sT}\, (\bm q_\sT^{2})^\alpha\, \mathcal{C}[w_{H}\, h_{1}^{\perp g}\, h_{1}^{\perp g}]=0$ for $\alpha=0,1$, model independently.
The case of $\alpha=0$ implies that linearly polarized gluons do not affect the $\bm q_\sT$-integrated cross section. This means that if one normalizes the differential cross sections to the integrated ones, the uncertainty from the hadronic matrix elements $\langle 0\vert {\cal{O}}_1 \vert 0\rangle$ cancels out. The case of  $\alpha=1$ implies that $R$ must have at least two nodes in $\bm q_\sT$, i.e.\ it has to flip sign at least twice, with\footnote{If $h_1^{\perp \, g}$ is of definite sign and thus has no nodes itself, then $R(\bm q_\sT^2=0) > 0$.} $R(\bm q_\sT^2=0) \neq 0$. This distinctive double node feature can be used to experimentally demonstrate the presence of the linear polarization 
of the gluons. 

It should be noted that the sign difference in the $R(\bm q_\sT^2)$ term in Eqs.\ \eqref{eq:CSeta} and \eqref{eq:CSchi0}, for different parities of the quarkonia, is completely analogous to the case of pseudoscalar and scalar Higgs boson production, respectively, considered in Ref.\ \cite{Boer:2011kf}. There is a clear difference between the quarkonium and Higgs production cases though. The existence of a scalar Higgs boson does not automatically imply the existence of a pseudoscalar Higgs boson (e.g.\ from a supersymmetric extension of the Standard Model\footnote{The recently observed Higgs boson signal at ATLAS and CMS cannot be accounted for by the $CP$-odd Higgs boson of the MSSM or NMSSM \cite{Benbrik:2012rm}.}), and even if it does exist, it is far from given that it couples to Standard Model particles in such a way that this sign change could be observable. In contrast, both types of charmonium and bottomonium states follow from the same theory and have been established experimentally, thus offering a definite possibility to cross check the predicted features, i.e.\ the double node modulation of the cross section {\it and} the sign difference. 

 The $\chi_{Q2}$ cross section does not receive a contribution from linearly polarized gluons, as it would require a helicity flip of 4 units, which is heavily suppressed. As a consequence, to leading order in $v^2$ the ratio of  $\chi_{Q0}$ to $\chi_{Q2}$ cross section yields a direct probe of the ratio $R(\bm q_\sT^2)$, as it simply equals $\frac{3}{4} (1+R(\bm q_\sT^2))$. Both the uncertainty from the hadronic matrix elements $\langle 0\vert {\cal{O}}_1 \vert 0\rangle$ [as follows from Eq.\  (\ref{mechi})] and from the unpolarized gluon TMD cancel out in this ratio. For the Higgs boson case, no analogous ratio can be formed. Another difference that may turn out to be crucial is due to the scale dependence. The contribution from linearly polarized gluons is expected to decrease with increasing scales, just as other TMD effects, such as the Sivers single spin asymmetry \cite{Aybat:2011ge,Aybat:2011ta,Anselmino:2012aa}. Therefore, for quarkonia with mass of order 3 or 10 GeV, the effect of linearly polarized gluons is expected to be larger than at a Higgs mass scale of 125 GeV. Of course, if both quarkonium and Higgs production can be studied, then that would allow for a valuable check of the scale dependence expected from TMD factorization. 

\begin{figure}[t]
\psfig{file=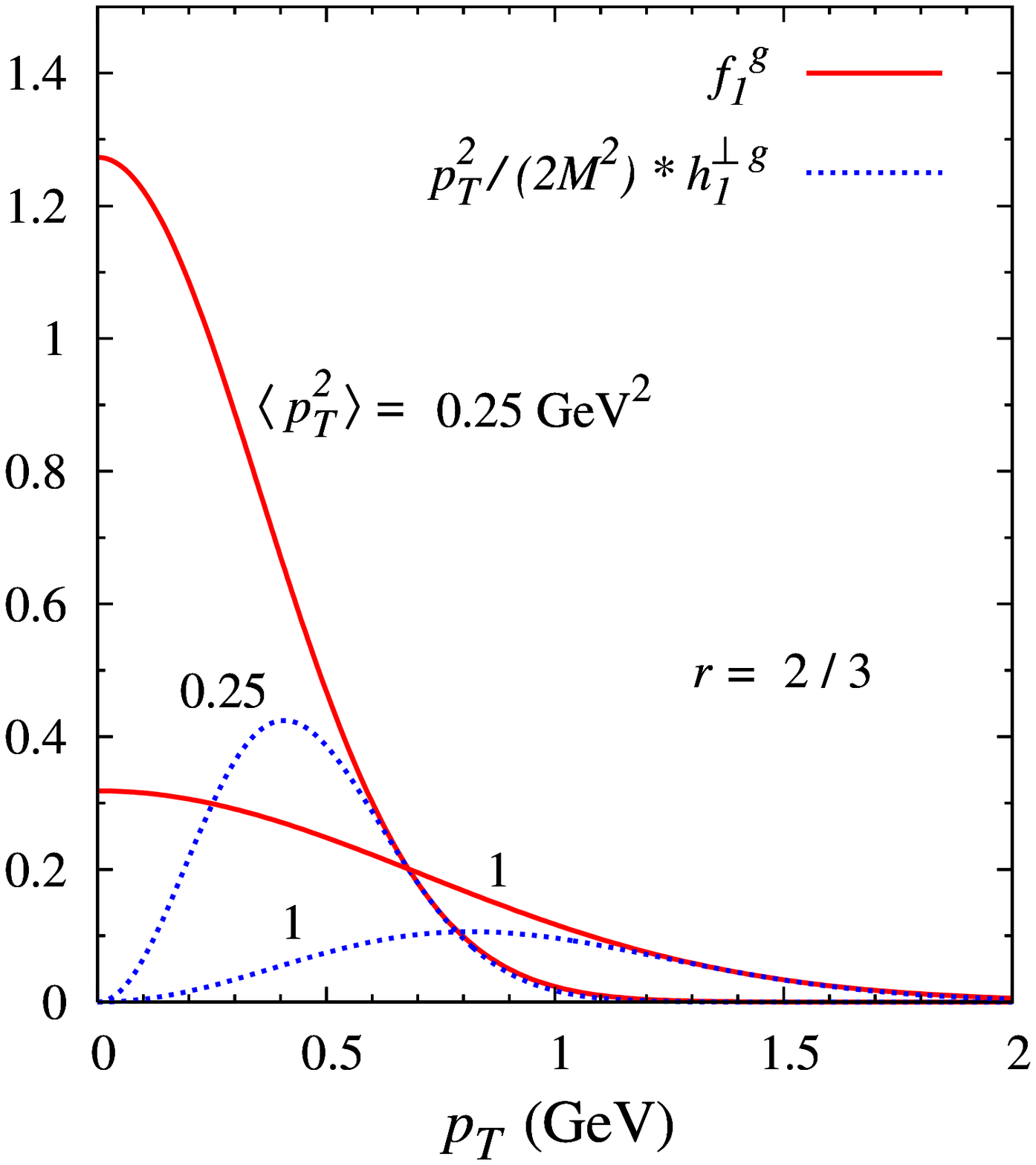, width=0.4\textwidth}
\psfig{file=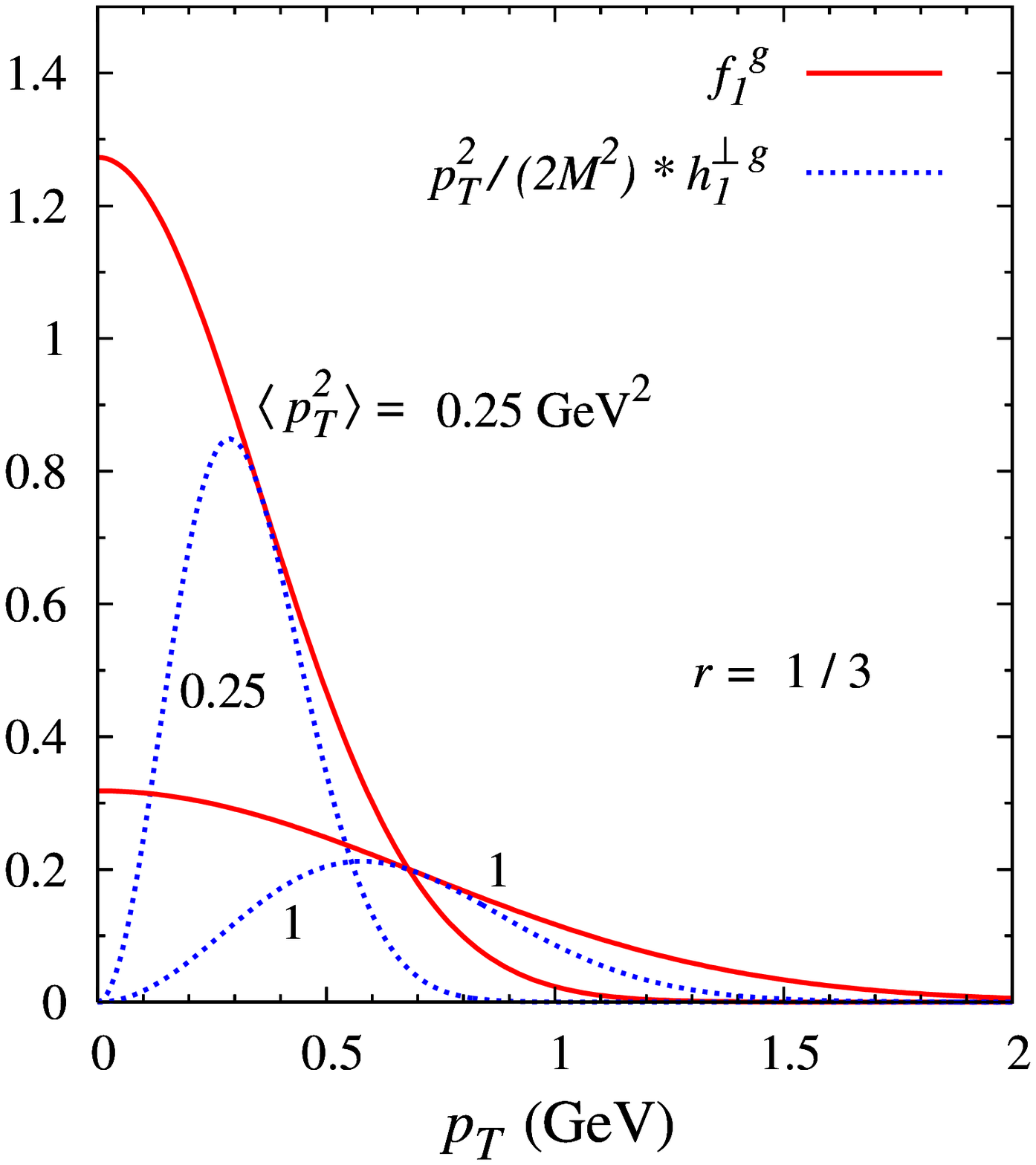, width=0.4\textwidth}
\caption{\it Gaussian distributions for $f_1^g$ and $h_1^{\perp \, g}$ [divided by $f_1^g(x)$] in units of GeV$^{-2}$ as functions of $p_\sT$ for two different
values of $\langle p_\sT^2 \rangle$ with $r=2/3$ (left) and $r=1/3$ (right).} 
\label{fig:input}
\end{figure}

It may be interesting to compare these findings to the expectations for the production cross sections in the color evaporation model (CEM)\footnote{For a detailed comparison between the CEM and NRQCD cf.\ \cite{Bodwin:2005hm}.}. In the CEM the probability to form a physical (colorless) quarkonium state is assumed to be independent of the color of the $Q \bar{Q}$ pair, which means that $gg\to Q \bar{Q} [^{2S+1}L^{(8)}_J] $ is assumed to produce a colorless quarkonium state by means of soft gluon emissions that do not affect the momenta or angular momenta. The cross sections for $gg\to Q\bar Q[^1 S_0^{(8)}]$,  $gg\to Q\bar Q[^3 P_0^{(8)}]$,  and $gg\to Q\bar Q[^3 P_2^{(8)}]$, then follow the same pattern as in Eqs.\ (\ref{eq:CSeta})-(\ref{eq:CSchi2}), namely $R$ enters with a positive (negative) sign for a $P$-even (odd) bound state, while $R=0$ when $J=2$. When integrated over $\bm q_\sT$ the results for these octet contributions, agree with the ones published in Ref.\ \cite{Cho:1995ce}. Therefore, we expect inclusion of color-octet contributions not to affect our conclusions: they are suppressed in the NRQCD based calculation and in the CEM they yield the same results as in the CSM. In the CEM sometimes the probability is also assumed to be independent of the spin of the $Q \bar{Q}$ pair \cite{Amundson:1996qr}. In that case no difference between the various $\chi_{QJ}$ states should exist. 
If $R(\bm{q}_\sT^2)$ is found to be nonzero in $\chi_{Q0}$, this version of the CEM would be ruled out.    

If the initial $gg$ pair is in a color singlet, the resulting $Q\bar{Q}$ pair is necessarily in a $C=+$ state \cite{Novikov:1977dq}, but for the color-octet case
it can also yield $C=-$ states. This means that in the CEM also $C=-$ quarkonium states can be produced in the $2 \to 1$ scattering process $gg\to Q \bar{Q}[^{2S+1}L^{(8)}_J] $. However, for $J=1$, such as for $J/\psi$ or $\Upsilon$ production, there is no contribution from $gg\to Q \bar{Q}[^{3}S^{(8)}_1] $ because of the Landau-Yang theorem, like for $\chi_{Q1}$. Therefore, also from CEM considerations we conclude that $J/\psi, \psi(2S)$ or $\Upsilon$ production will not be useful for studies of $h_1^{\perp\, g}$ at small transverse momentum.

At order $\alpha_s^3$ $1^{\pm\pm}$ states can be produced through the subprocess $gg\to Q \bar{Q} g$, but the contribution from $h_1^{\perp\,g}$ to the cross section is found to be power suppressed in  these and all other quarkonium cases. Therefore, we do not expect any unsuppressed contribution from $h_1^{\perp \, g}$ in $C=\pm$ quarkonium production at large transverse momentum in hadronic collisions. Of course it may be possible that for example $g g \to \chi_{c0} \to J/\psi \, \gamma$ produces a contribution for $J/\psi$'s with  larger transverse momentum, but this will only be significant on the intermediate $J=0$ resonance which itself has small transverse momentum. Such cases simply follow the pattern of the results obtained in this paper. 
Note that in this example the final state photon (or gluon jet) momentum 
needs to be measured explicitly, otherwise no unsuppressed contribution from $h_1^{\perp \, g}$ arises.
  
\section{Numerical results}

\begin{figure}[t]
\psfig{file=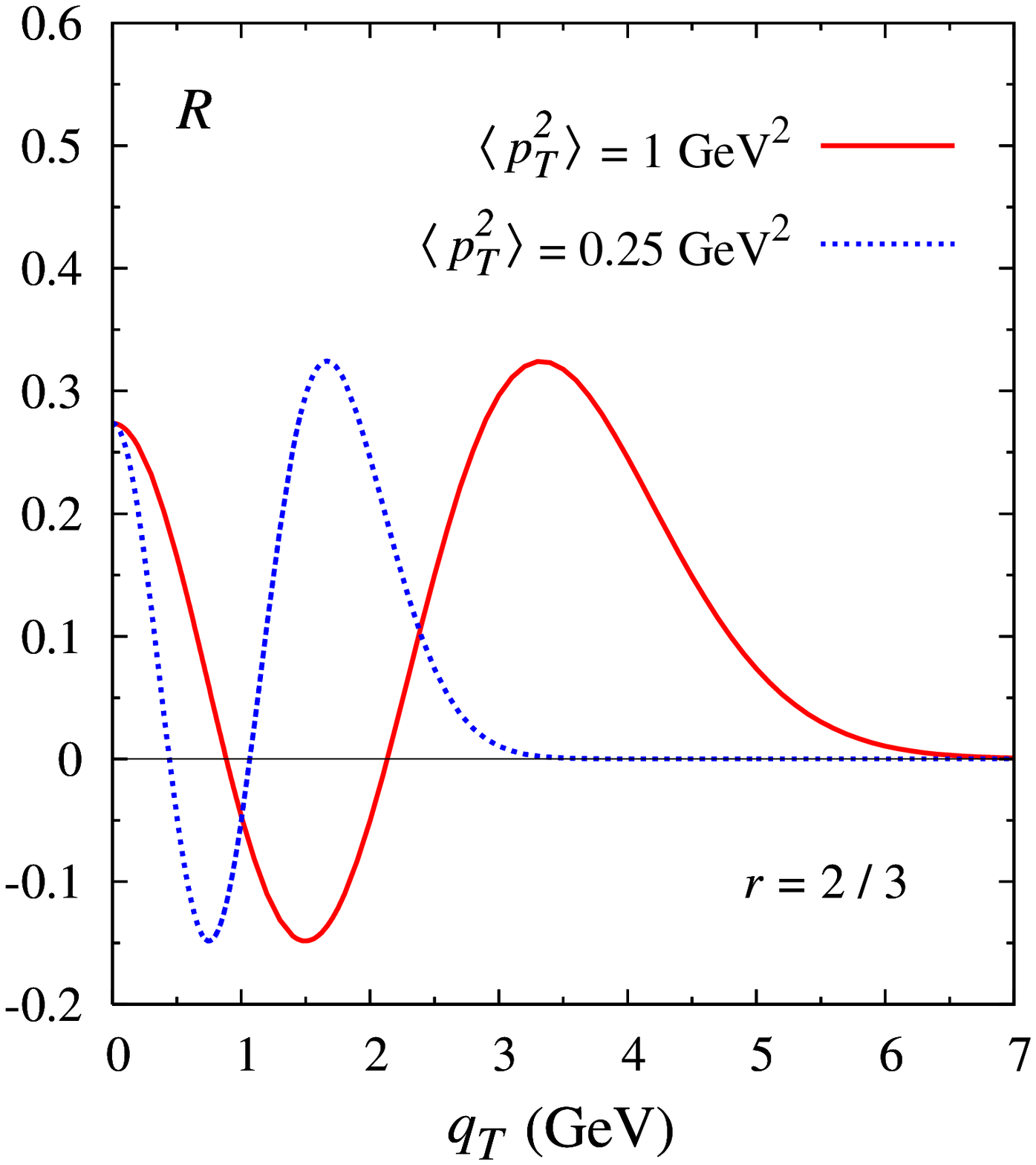, width=0.4\textwidth}
\psfig{file=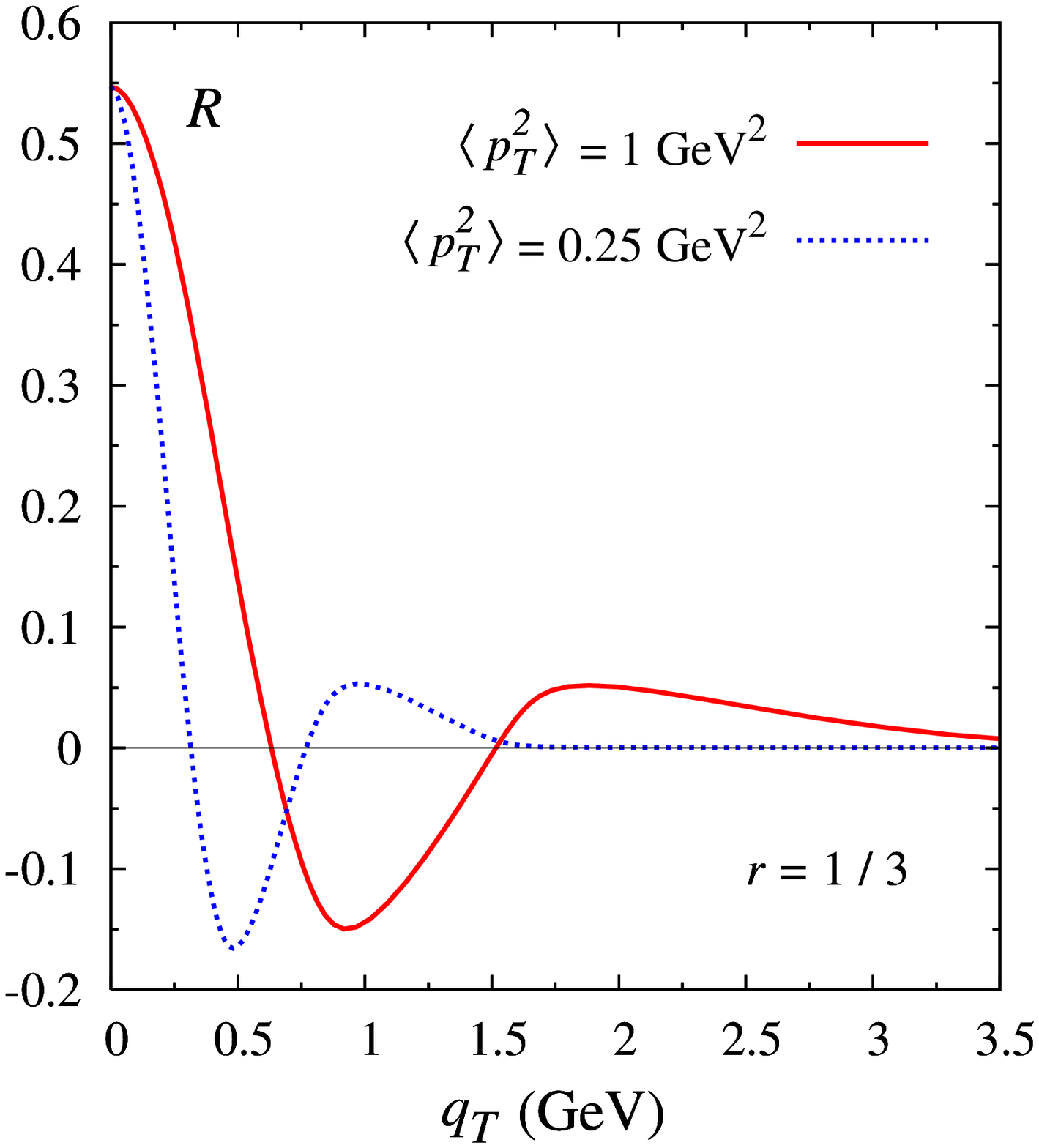, width=0.4\textwidth}
\caption{\it Ratio $R(r, \bm q_\sT^2) =\mathcal{C}[w\,h_1^{\perp g}\,
h_1^{\perp  g}]/\mathcal{C}[f^g_1\,f_1^g]$ as a function of $q_\sT$ for two different values of the input parameter $r$.}
\label{fig:R}
\end{figure}

In this section we estimate the possible size of the contribution from $h_1^{\perp \, g}$ to the quarkonium production cross section, calculated at order $\alpha_s^2$ within the framework of the color-singlet model. 
The effect will be maximal when the unknown function $h_1^{\perp g}$ saturates  
its model-independent positivity bound \cite{Mulders:2000sh},
\begin{equation}
\frac{\bm p_\sT^2}{2M_h^2}\,|h_1^{\perp g}(x,\bm p_\sT^2)|\le f_1^g(x,\bm p_\sT^2)\,,\label{eq:Bound}
\end{equation}
which is valid for all values of $x$ and $\bm p_\sT$. 
Adopting a standard approach, we assume that the gluon TMDs have a 
simple Gaussian dependence on transverse momentum \cite{Boer:2011kf},
\begin{equation}
f_1^g(x,\bm p_\sT^2) = \frac{f_1^g(x)}{\pi \langle  p_\sT^2 \rangle}\,
\exp\left(-\frac{\bm p_\sT^2}{\langle  p_\sT^2 \rangle}\right)\,,\label{eq:Gaussf1}
\end{equation}
where $f_1^g(x)$ is the collinear gluon distribution and the width
$\langle p_\sT^2 \rangle$ will in general depend on the energy scale, which is set by the quarkonium mass $M$. Moreover $\langle p_\sT^2 \rangle$ is taken to be independent of $x$.
The bound in Eq.\ (\ref{eq:Bound}) is satisfied (but not everywhere saturated) by the form
\begin{equation}
h_1^{\perp g}(x,\bm p_\sT^2)=\frac{M^2 f_1^g(x)}{\pi \langle p_\sT^2 \rangle^2}\frac{2(1-r)}{r}\,
\exp\left(1 -\frac{1}{r}\,\frac{\bm p_\sT^2}{\langle p_\sT^2 \rangle}\right)\,.\label{eq:Gaussh1perp}
\end{equation}
In Fig.\ \ref{fig:input} the $\bm p_\sT$ dependence of $f_1^g$ and $h_1^{\perp g}$ is shown 
for two values of the input parameter $r$ ($r=2/3$ and $r=1/3$) and two  values of the Gaussian width: $\langle p_\sT^2 \rangle = 0.25\,\mathrm{GeV}^2$ and $\langle p_\sT^2 \rangle = 1\,\mathrm{GeV}^2$, which are considered relevant for charmonium and bottomonium, respectively. Recall that the   expressions given in the previous section apply for $q_\sT^2 \ll M^2$. This means that for $q_\sT^2 \sim M^2$ we expect the approach and in particular the Gaussian falloff to be unrealistic, rather one expects a power-law falloff of $h_1^{\perp \, g}$ that would result from a perturbative collinear treatment \cite{Sun:2011iw}. In practice, this difference may not matter much, as the cross section and the modulation from linearly polarized gluons are expected to be quite small already in this tail region. 

\begin{figure}[t]
\psfig{file=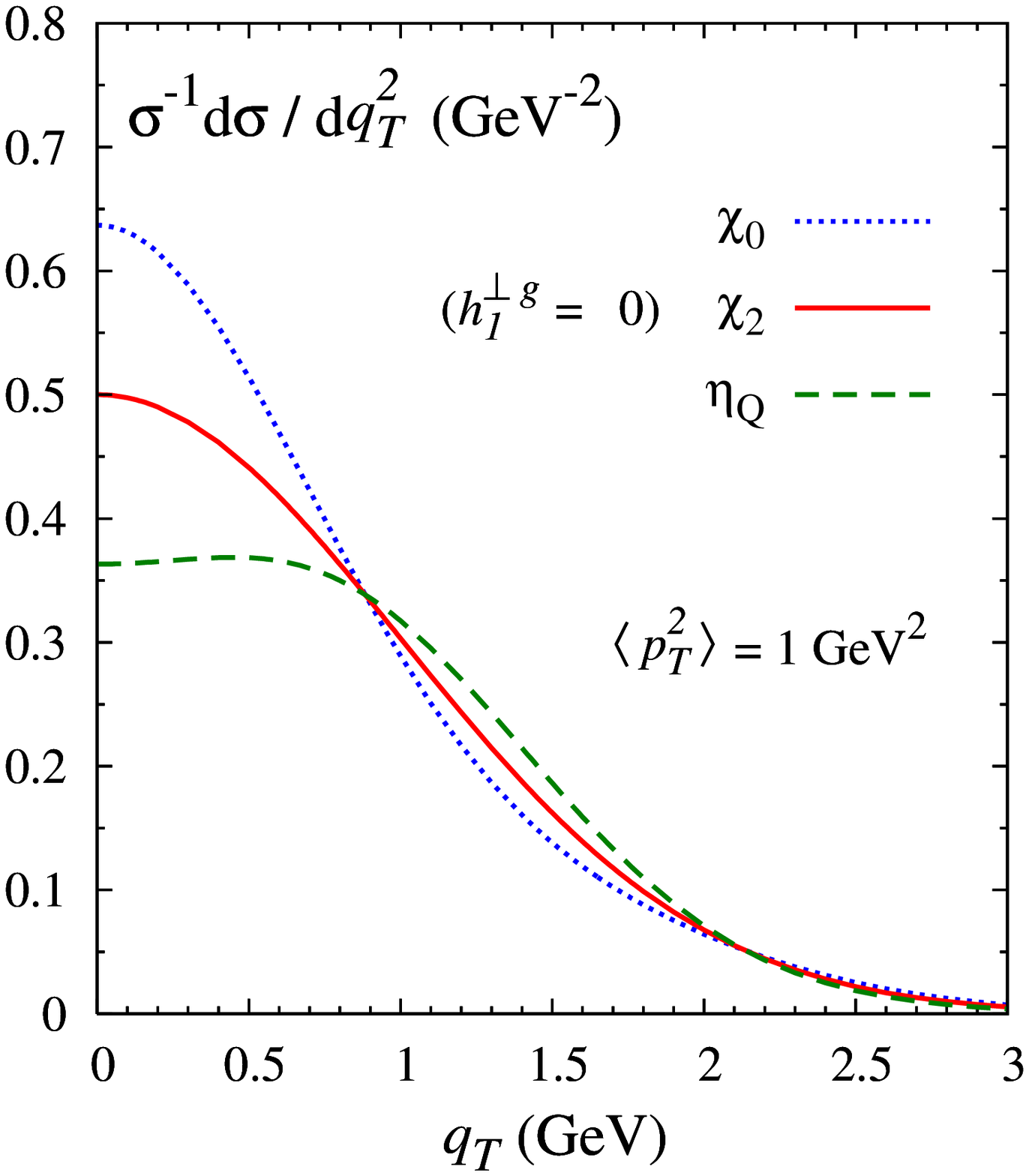, width=0.4\textwidth}
\psfig{file=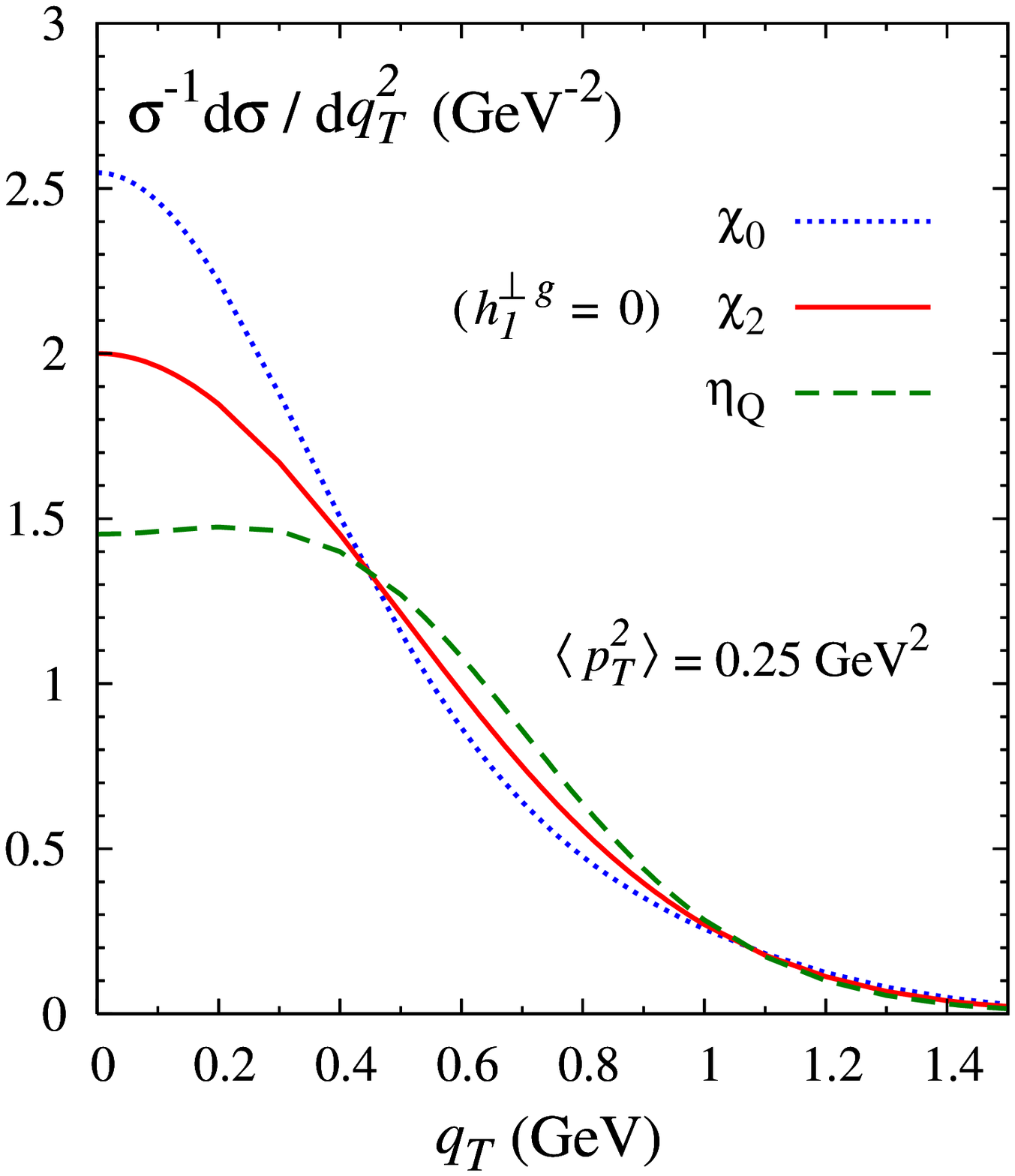, width=0.4\textwidth}
\caption{\it Transverse momentum distributions for the different quarkonia, using the distribution functions in Eqs.\ (\ref{eq:Gaussf1}) and  (\ref{eq:Gaussh1perp}) with $r=2/3$, for two different values of   $\langle p_\sT^2 \rangle$. }
\label{fig:TMD}
\end{figure}
\begin{figure}[b]
\psfig{file=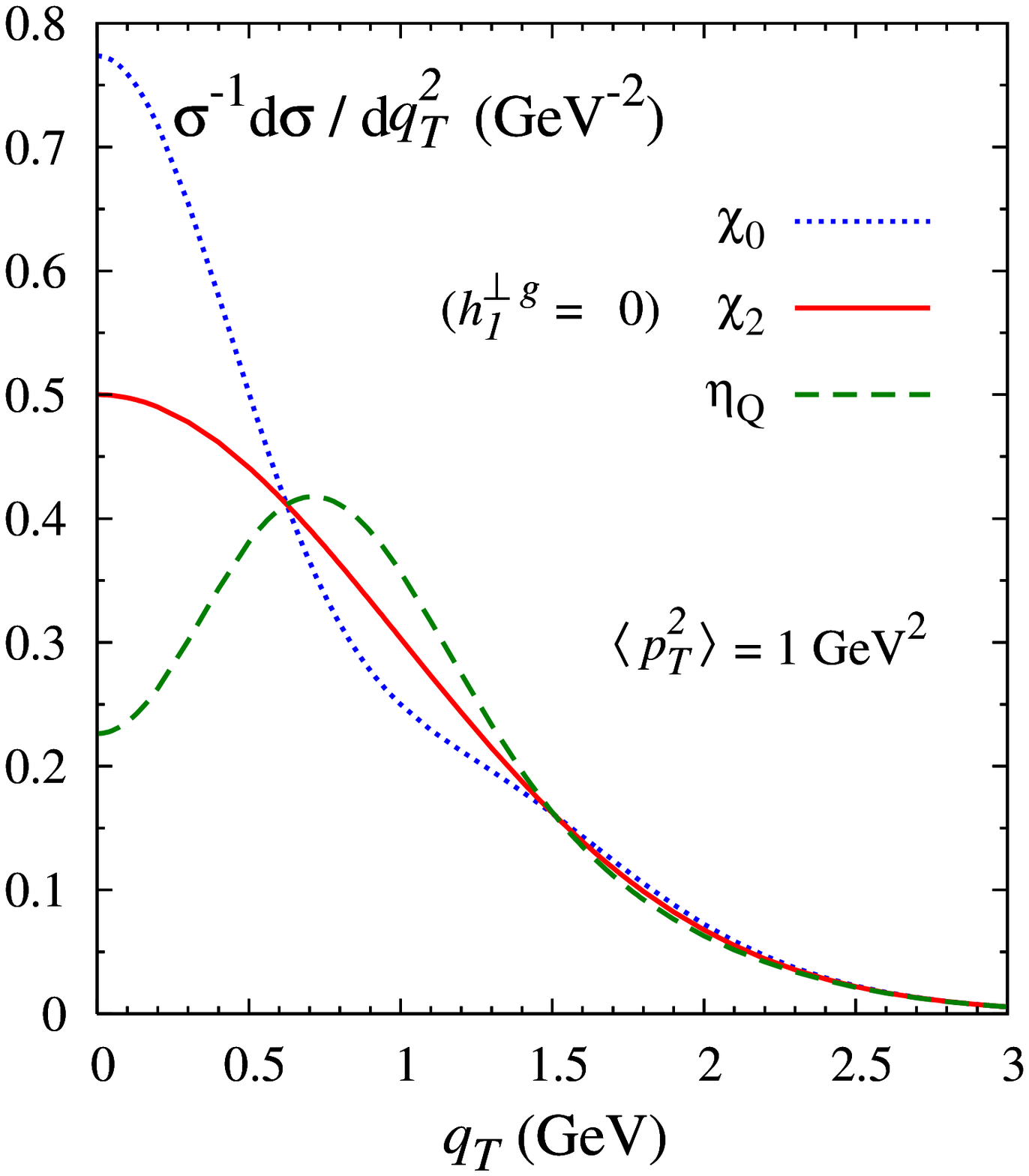, width=0.4\textwidth}
\psfig{file=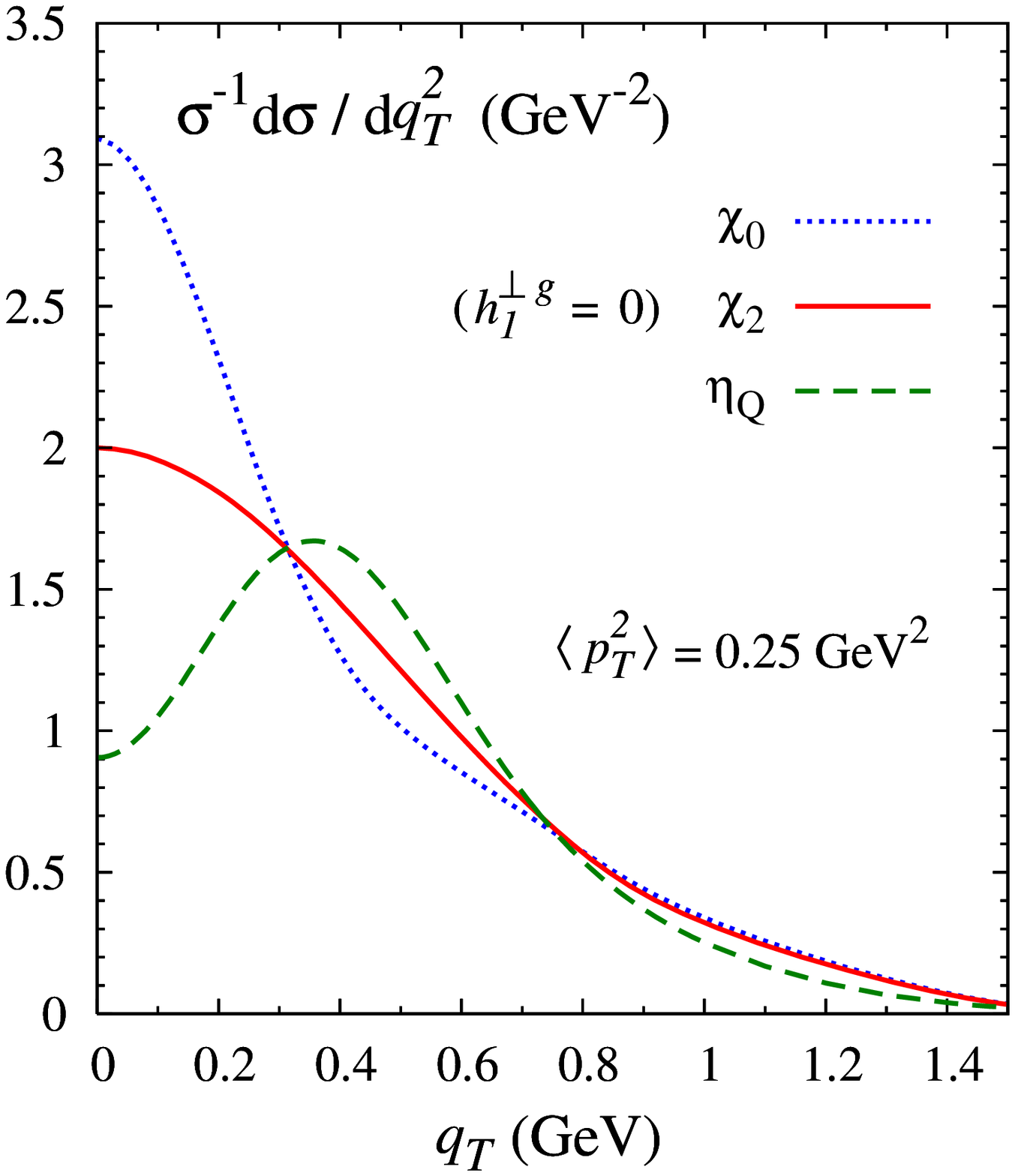, width=0.4\textwidth}
\caption{\it Same as in Fig.\ \ref{fig:TMD}, but with $r=1/3$.}  
\label{fig:TMD_r033}
\end{figure}

The convolution integrals that appear in Eqs.\  (\ref{eq:CSeta})-(\ref{eq:CSchi2}) can be evaluated analytically: 
\begin{eqnarray} 
 \mathcal{C}[f^g_1\,f_1^g] & = & \frac{f_1^g(x_a)\,f_1^g(x_b)}{2\pi \langle p_\sT^2 \rangle}\,\mathrm{exp}
\left(-\frac{1}{2}\,\frac{\bm q_\sT^2}{\langle p_\sT^2 \rangle}\right),\label{eq:GaussConvf1}\\ 
 \mathcal{C}[w\,h_1^{\perp g}\,h_1^{\perp g}]  & = & \frac{f_1^g(x_a)\,f_1^g(x_b)}{4 \pi 
\langle p_\sT^2 \rangle}\,r(1-r)^2\left[1 -\frac{1}{r}\,\frac{\bm q_\sT^2}{\langle p_\sT^2 \rangle}
 +\frac{1}{8r^2}\,\frac{(\bm q_\sT^2)^2}{\langle p_\sT^2 \rangle^2}\right]\,\mathrm{exp}\left(2- \frac{1}{2 r}\,\frac{ \bm q_\sT^2}{
 \langle p_\sT^2 \rangle}\right)\,.\label{eq:GaussConvh1perp}
\end{eqnarray}
Their ratio,
\begin{equation}
R(r, \bm q_\sT^2) = \frac{1}{2}\,r\,(1-r)^2 \left[1 -\frac{1}{r}\,\frac{\bm q_\sT^2}{\langle p_\sT^2 \rangle}
 +\frac{1}{8r^2}\,\frac{(\bm q_\sT^2)^2}{\langle p_\sT^2 \rangle^2}\right]\,\mathrm{exp}\left(2- \frac{1-r}{2 r}\,\frac{ \bm q_\sT^2}{
 \langle p_\sT^2 \rangle}\right)\,,
\end{equation} 
measures the relative size of the contribution by linearly polarized gluons. The behavior of $R$ as a function of $q_\sT$, presented in Fig.\ \ref{fig:R} for $r=2/3$ (left panel) and for  $r=1/3$ (right panel), is characterized by the expected double node structure of $\mathcal{C}[w  \,h_1^{\perp g} \,h_1^{\perp g}]$ \cite{Boer:2011kf}.

The transverse momentum distributions of $\eta_Q$ and $\chi_{Q0,2}$ are therefore given by
\begin{eqnarray}
\frac{1}{\sigma(\eta_Q)} \, \frac{\d\sigma (\eta_Q)}{\d \bm q_\sT^2} & = & \left [1-R(r, \bm q_\sT^2)\right ] \, \frac{1}{2 \langle p_\sT^2 \rangle}  \, {\rm exp} \left (- \frac{\bm q_\sT^2}{2 \langle p_\sT^2 \rangle }\right ),  \\
\frac{1}{\sigma(\chi_Q)} \, \frac{\d\sigma (\chi_{Q 0})}{\d \bm q_\sT^2} & = &  \left [1+R(r, \bm q_\sT^2)\right ] \, \frac{1}{2 \langle p_\sT^2 \rangle} \, {\rm exp} \left (- \frac{\bm q_\sT^2}{2 \langle p_\sT^2 \rangle }\right ),\\
\frac{1}{\sigma(\chi_Q)} \, \frac{\d\sigma (\chi_{Q 2})}{\d \bm q_\sT^2} & = &  \frac{1}{2 \langle p_\sT^2 \rangle} \, {\rm exp} \left (- \frac{\bm q_\sT^2}{2 \langle p_\sT^2 \rangle} \right )\,,
\end{eqnarray}
and they are shown in Figs.\ \ref{fig:TMD} and \ref{fig:TMD_r033} for different values of the input parameter $r$ and the Gaussian
 width $\langle p_\sT^2\rangle$. 
As mentioned, the distributions for $\eta_{c,b}$ and 
$\chi_{c,b\,0}$ are similar to the ones for a pseudoscalar and a scalar Higgs boson, cf.\ \cite{Dunnen:2012ym}. Unfortunately, there are no data available yet  to fit or bound the modulation, neither for charmonium nor for bottomonium. Hopefully, LHCb will be able to provide such data in the near future. Existing data for $\chi_c$ production (\cite{Kourkoumelis:1978mj} and more recently, from Tevatron \cite{Abe:1997yz} and LHC \cite{LHCb:2012af}) is from $\chi_c \to J/\psi\; \gamma$, which is predominantly $\chi_{c1}$ and $\chi_{c2}$. Other decay channels, with a larger $\chi_{c0}$ component, should be considered instead.   

To obtain numerical estimates for the production cross sections, we have employed a Gaussian model. Since the $x$ values of the gluons in one of the two protons can be rather small at LHC, one may also consider to calculate the cross sections within a small-$x$ formalism. It has been shown that the distribution of linearly polarized gluons in various small-$x$ treatments, the color dipole model \cite{Metz:2011wb,Dominguez:2011br} and a color glass condensate model \cite{Metz:2011wb,Schafer:2012yx}, is quite sizable, in fact, can even saturate its bound. Therefore, we do not expect small-$x$ effects to necessarily lead to smaller modulations compared to those obtained here for the Gaussian model.

\section{Summary and conclusions}
In this paper $C=+$ quarkonium production in hadronic collisions has been studied within the framework of TMD factorization in combination with the nonrelativistic QCD based color-singlet quarkonium model. It is shown that the transverse momentum spectra of   scalar ($\chi_{c0}, \chi_{b0}$) and pseudoscalar quarkonium ($\eta_c, \eta_b$) production are modified in different ways by linearly polarized gluons that can be present inside {\it unpolarized} hadrons. The scalar spectra receive a modulation $R(\bm q_\sT^2)$, whereas for pseudoscalars it is of opposite sign, $-R(\bm q_\sT^2)$. The modulation itself displays a characteristic double node structure, the magnitude of which is not yet known however. The advantage of using (pseudo)scalar final states is that no angular analysis needs to be performed to probe the effects of the linear polarization of gluons. Their effects on the production of higher angular momentum quarkonium states are strongly suppressed. As a consequence, comparison of $\chi_{Q0}$ and $\chi_{Q2}$ production can help to cancel out uncertainties. Similar results are not obtained for $C=-$ states, as they turn out not to be sensitive to the linear polarization at leading order. We thus conclude that $C=+$ charmonium and bottomonium states offer an excellent way to access and extract the distribution of linearly polarized gluons inside unpolarized hadrons and to check the predicted features: the double node structure and the sign difference between scalar and pseudoscalar production. Together with the analogous study in Higgs production at LHC, quarkonium production can moreover be used to test the scale dependence of the linearly polarized gluon distribution over a large energy range. Experimental opportunities to study these new proposals are currently offered by LHCb and the proposed fixed-target experiment AFTER at LHC.

\begin{acknowledgments}
We are grateful to Stan Brodsky, Jean-Philippe Lansberg and Francesco Murgia
 for useful discussions and feedback. C.P.\ acknow\-ledges financial support 
from the European Community under the FP7 ``Capacities - Research Infrastructures'' program (HadronPhysics3, Grant Agreement 283286).
\end{acknowledgments}


\end{document}